\documentclass[12pt,preprint]{aastex}
\begin{document}
\title{
OBSERVATIONAL SEARCH FOR PeV--EeV TAU NEUTRINO \\ 
FROM GRB081203A
} 
\author{Y.~Aita\altaffilmark{1},
T.~Aoki\altaffilmark{1},
Y.~Asaoka\altaffilmark{1},
T.~Chonan\altaffilmark{1},
M.~Jobashi\altaffilmark{1},
M.~Masuda\altaffilmark{1},
Y.~Morimoto\altaffilmark{1},
K.~Noda\altaffilmark{1},
M.~Sasaki\altaffilmark{1},
J.~Asoh\altaffilmark{2},
N.~Ishikawa\altaffilmark{2},
S.~Ogawa\altaffilmark{2},
J.~G.~Learned\altaffilmark{3},
S.~Matsuno\altaffilmark{3},
S.~Olsen\altaffilmark{3},
P.~-M.~Binder\altaffilmark{4},
J.~Hamilton\altaffilmark{4},
N.~Sugiyama\altaffilmark{5},
and
Y.~Watanabe\altaffilmark{6}\\
(Ashra-1 Collaboration)}
\altaffiltext{1}{Institute for Cosmic Ray Research, University of Tokyo, Kashiwa, Chiba 277-8582, Japan; \\ 
\hspace*{0.7cm} asaoka@icrr.u-tokyo.ac.jp, sasakim@icrr.u-tokyo.ac.jp}
\altaffiltext{2}{Department of Physics, Toho University, Funabashi, Chiba 274-8510, Japan} 
\altaffiltext{3}{Department of Physics and Astronomy, University of Hawaii at Manoa, Honolulu, HI 96822, USA}
\altaffiltext{4}{Department of Physics and Astronomy, University of Hawaii at Hilo, Hilo, HI 96720-4091, USA}
\altaffiltext{5}{Department of Physics and Astrophysics, Nagoya University, Nagoya, Aichi 464-8601, Japan} 
\altaffiltext{6}{Department of Engineering, Kanagawa University, Yokohama, Kanagawa 221-8686, Japan} 

\begin{abstract}
We report the first observational search for tau neutrinos ($\nu_{\tau}$) 
from gamma-ray bursts (GRBs) using one of the Ashra light collectors.  
The earth-skimming $\nu_{\tau}$ technique of imaging Cherenkov $\tau$ 
showers was applied as a detection method. 
We set stringent upper limits on the $\nu_{\tau}$ fluence in PeV--EeV 
region for 3780 s (between 2.83 and 1.78 hr before) and another 
3780 s (between 21.2 and 22.2 hr after) surrounding GRB081203A triggered by the {\it Swift} satellite.  
This first search for PeV--EeV $\nu_{\tau}$ complements other experiments 
in energy range and methodology, and suggests the prologue of ``multi-particle astronomy'' 
with a precise determination of time and location.
\end{abstract}

\section{Introduction}
Gamma-ray bursts (GRBs) eject the most energetic outflows  
in the observed universe,
with jets of material 
expanding relativistically into the surrounding interstellar matter
with a Lorentz factor $\Gamma$
of 100 or more.
Energy dissipation processes involving
nonthermal interactions between particles 
are thought to play an important role
in GRBs, but remain observationally unresolved.
The detection of PeV--EeV~ neutrinos ($\nu$s) 
from a GRB provides direct evidence for 
the acceleration of hadrons into the EeV range,
and of photo-pion interactions in the GRB.
The GRB standard model 
\citep[][and references therein]{Meszaros06}, 
which is based on internal/external shock acceleration, 
has been used to describe the general features of a GRB and the 
observed multi-wavelength afterglow. 
However, the standard model cannot well reproduce recent observational results.
The early X-ray afterglows detected by {\it Swift} 
exhibited a canonical behavior of steep-flat-steep in their light curve
\citep{Nousek06}. 
In 10\%--15\% of GRBs, precursor activities were observed 
\citep{Burlon08}. 
In some cases, the precursor preceded the main burst by several hundred
seconds with significant energy emission. 
In the {\it Fermi} observations of GRB090510 and GRB090902B, 
spectral fits revealed a hard power-law component 
\citep{Abdo09b}. 
The {\it Swift} and {\it Fermi} observations
of GRB090510 detected gamma rays ($\gamma$s) in the GeV
range up to 200~s after the lower energy trigger 
\citep{DePasquale10}. 
Although many authors have proposed theoretical models to reproduce
the complicated time evolution of  
GRBs
and the high energy components in the 
prompt emission, 
none of these models are conclusive 
\citep{Ackermann10}. 
To better understand the ambiguous mechanisms of GRBs, 
observational probes of the
optically thick region of the electromagnetic components, 
as well as hadron acceleration processes 
throughout the precursor, prompt, and afterglow phases are required. 
Very high energy (VHE) $\nu$s can be used as direct observational probes, 
which are effective even in 
optically thick regions.
A monitor search with sufficient time and spatial resolution
and survey capability for VHE$\nu$s associated with GRBs is plausible. 

The earth-skimming tau neutrino ($\nu_{\tau}$) 
technique, 
which detects extensive air showers 
\citep{Fargion02}, 
has the advantage of a large target mass,
since it uses air showers produced by decay particles of tau leptons ($\tau$s)  
in the atmosphere as the observed signals.  
$\tau$s emerge out of the side of the mountain or the ground facing the detector; 
they are the product of interactions between VHE $\nu_{\tau}$ and 
the earth matter they traverse. 
Above 1~EeV, air fluorescence observations based on the
earth-skimming $\nu_{\tau}$ technique have been reported 
\citep{Abraham08}. 
No air Cherenkov observation was made to date 
based on the earth-skimming $\nu_{\tau}$ technique with air showers induced by $\tau$ decays 
(hereafter referred to as the Cherenkov $\tau$ shower method). 
But it can achieve sufficient detection sensitivity in the PeV--EeV
region to be useful in the search for $\nu$s originating from
hadrons accelerated to EeV at astronomical objects. 
Additional advantages of the Cherenkov $\tau$ shower method 
are its perfect shielding of cosmic-ray secondary particles, 
highly precise arrival direction determination 
for primary $\nu_{\tau}$ and 
negligible background contamination by atmospheric $\nu$s
in the PeV--EeV energy range.

\section{Ashra Experiment and Observation}
The all-sky survey high-resolution air-shower detector (Ashra) 
is a complex of unconventional optical collectors 
that image VHE air showers in a 42$^\circ$ diameter field of view (FOV) 
covering 77\% of the entire night sky with a resolution of 
a few arcminutes
\citep{Sasaki08,Aita07,Sasaki07}. 
The first phase of the Ashra experiment (Ashra-1) 
was constructed on Mauna Loa at 3300~m above sea level on Hawaii Island,  
and includes an observatory.
Ashra-1 uses
electrostatic lenses \citep{AsaokaSasaki11} in addition to an optical system
to generate convergent beams, 
enabling a very low cost and high performance image sensor, 
providing a high resolution over a wide FOV. 
The electron optics use an image pipeline to transport the image
from the focal sphere of the reflective mirror optical system. 
After the light from the image is split,
it is transported
to both a trigger device and
high-gain, high-resolution 
complementary metal-oxide semiconductor image sensor. 

\begin{figure}[bt!]
\begin{center}
\includegraphics[width=0.8\hsize]{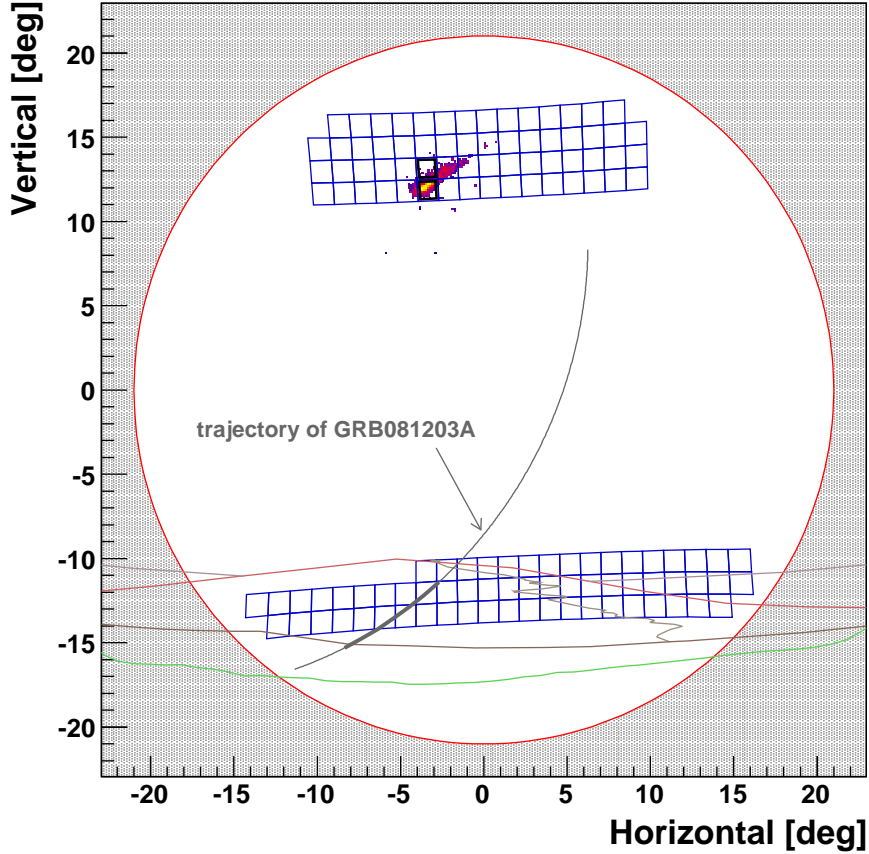} 
\vspace*{-4mm}
\end{center}
\caption{\label{fig:evdisp}
Boundary 
(large red circle) 
between the inside (open circle) and outside (hatched area) of the FOV
of the Ashra-1 light collector, which faces Mauna Kea,
and the layout of trigger pixel FOVs 
(blue boxes) 
for Cherenkov $\tau$ shower observation. 
Repositioned array of the trigger pixel FOVs 
(upper blue boxes) to check the detection sensitivity 
with ordinary cosmic-ray air showers at a higher elevation. 
Firing trigger pixels (thick blue boxes) of 
an observed image of a cosmic-ray air shower readout along the trigger 
(points).
An extended portion of the trajectory of GRB081203A counterpart 
(circular arc),
the segment of this trajectory
used in the $\nu_{\tau}$ search (thick circular arc),
the ridge lines of Mauna Kea (red) and Mauna Loa (green) mountains, 
the horizon, and Mauna Kea access road are shown.
}
\end{figure}

One of the Ashra light collectors built on Mauna Loa
has the geometrical advantages of 
not only
facing Mauna Kea, allowing it
to encompass the large target mass of Mauna Kea in the observational FOV,
but has also an 
appropriate distance of $\sim$30~km from Mauna Kea, 
yielding good observational efficiency when imaging 
air-shower Cherenkov lights which are directional 
with respect to the air-shower axis. 
Using the advanced features,
we performed commissioning search for  
Cherenkov $\tau$ showers for 197.1 hr 
between October and December of 2008. 
We served limited 62 channels of photomultiplier tubes (PMTs) as trigger sensors 
prepared for the commissioning runs 
to cover the view of the surface area of Mauna Kea,
maximizing the trigger efficiency for Cherenkov $\tau$ showers 
from Monte Carlo (MC) study,
as shown in Fig.~\ref{fig:evdisp}. 
Adjacent-two logic was adopted to
trigger the fine imaging, 
by judging discriminated waveform signals from 
each pixel of the multi-PMT trigger sensor.
During the search period,
$\sim$2 hr before the trigger of GRB081203A 
\citep{GCN8595}, 
GRB counterpart 
(R.A. 15:32:07.58, decl. +63:31:14.9) 
passed behind Mauna Kea, as 
viewed from the Ashra-1 observatory.
Using the same light collector but continuously taking fine images
of split lights just before the trigger and readout sensors
in the image pipeline, 
we set a limit on the
light curve as a function of time in the region of 300~s bracketing 
the {\it Swift} trigger time for GRB081203A,
including the precursor and prompt afterglow optical 
counterpart
\citep{GCN8632}.

\section{Analysis}
To investigate the features, selection criteria, 
detection efficiency, and background rate 
for the observation of Cherenkov $\tau$ shower images, 
we generated $\nu_\tau$ MC events
with primary energies of 1~PeV to 100~EeV by 0.5 decade steps,
which entered into the rock of Mauna Kea 
uniformly and isotropically from a sufficiently large aperture.
We used a geodetic database around Hawaii island 
\citep{Mooney98},  
and surveyed for ourselves the position
of the observatory and the terrain of the mountain and 
its surroundings.
To study the generation and propagation of $\tau$s 
in the earth and in the mountain, 
we used PYTHIA 
\citep{PYTHIA6154} 
to simulate the charged current interaction of
$\nu_{\tau}$s with nucleons and 
GEANT4
\citep{Geant4} 
for the energy loss 
due to pair production and bremsstrahlung in $\tau$ propagation.
Photonuclear interaction was 
estimated using the differential crosssection
given in \cite{Dutta2001} and \cite{ALLM}. 
TAUOLA \citep{TAUOLA24} was used for $\tau$ decays and 
CORSIKA \citep[thinning=10$^{-7}$;][]{CORSIKA} for 
air showers induced by $\tau$ decays. 
To simulate the Ashra detector,
we took into account
the geometry,
light collection area,
mirror reflection,
corrector lens transmittance,
and the quantum efficiency of the photoelectric tubes,
so that the fine images corresponding 
to the trigger judgement were simulated in an event-by-event manner. 
$\Delta \theta_{\tau}$, 
the deflection angle of $\tau$
with respect to the primary $\nu_\tau$,
was estimated to be significantly less than 1~arcmin
at energies above 1~PeV
due to the physical processes 
occurring
in the rock of the earth and of the mountain 
(Y. Asaoka \& M. Sasaki 2011, to be published). 
In addition, the reconstructed $\tau$ shower axis
can point toward the $\nu_\tau$ object 
within an accuracy of $0^\circ\!\!.1$ 
if the resolution of the image is sufficiently high. 
Therefore, the
Cherenkov $\tau$ shower induced by PeV--EeV $\nu_\tau$ 
is a fine probe into VHE hadron accelerators
once an image is obtained with sufficient resolution. 
Such images can be obtained with the Ashra detector. 

The photometric and trigger sensitivity calibration 
of the Ashra light collector was based on 
a very stable YAP(YAlO$_{3}$:Ce)-light pulser \citep{YAP92} 
which was placed at the center of the input window of 
the photoelectric lens image tube 
mounted on the focal sphere of the optical system 
and illuminated it.
Non-uniformity in the detector gain 
due to the input light position  
was relatively corrected by mounting
a spherical plate uniformly covered with luminous paint
on the input window.
To correct for the time variation of 
the photometric and trigger sensitivity 
because of variations in atmospheric optical thickness,
which were mainly due to clouds and hazes during the observation period,
we performed careful cross-calibration to compare the instrumental 
photoelectric response with the photometry of standard stars
such as BD+75D325 of B-magnitude 9.2, for which the detected images
passed through the same optical and photoelectric instruments except
for the final trigger-controlled readout device.
We estimated the systematic uncertainty on the basis of our understanding of
the detector sensitivity
to be 30\%
after applying a combination of 
the above three complementary 
calibration procedures.

\begin{figure}[bt!]
\begin{center}
\includegraphics[width=0.85\hsize]{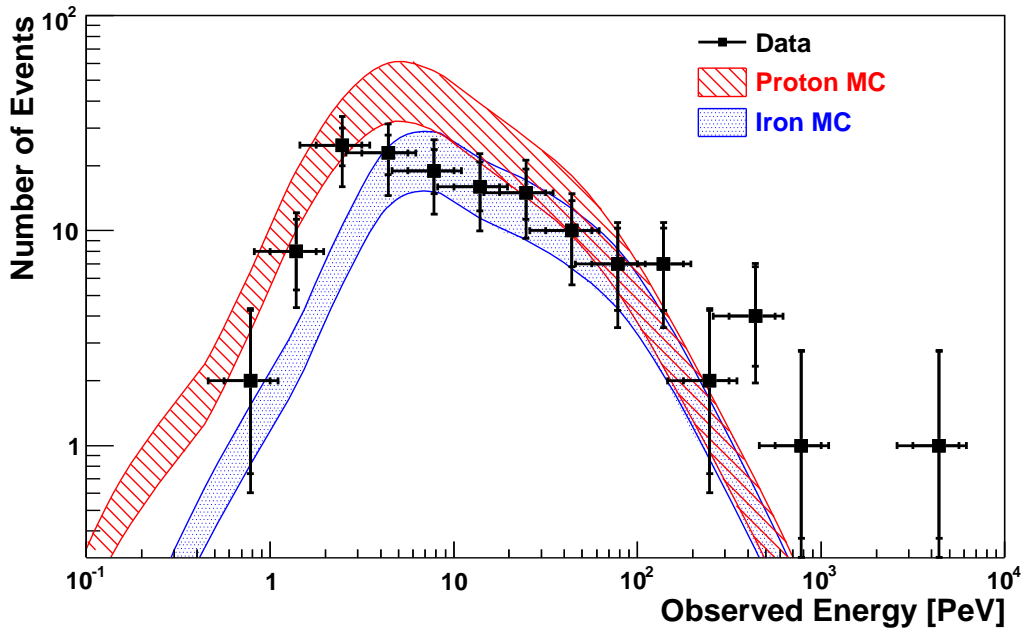} 
\end{center}
\caption{\label{fig:pedist}
Observed cosmic-ray flux spectrum 
(filled box)
with bars indicating statistical and systematic errors
and the MC predictions for
proton primary (hatched band) and iron primary (shaded band) 
assumptions.
The width of the bands shows the evaluated systematic error 
of 30\% of the MC prediction
(see the text). 
}
\end{figure}

To validate the
detection sensitivity and gain calibration 
for the Cherenkov $\tau$ shower, 
we detected and analyzed
140 events of ordinary cosmic-ray air-shower Cherenkov images for a total of 
44.4~hr in 2008 December using 
the same instruments used in the Ashra light collector, but
after rearranging the trigger pixel layout to view
the sky field above Mauna Kea (Fig.~\ref{fig:evdisp}). 
In the cosmic-ray observation, the trigger pixel layout is centered at 
zenith angle of $\sim$65$^\circ$. 
Due to the directionality of the air-shower Cherenkov lights,
the photometric detection of the lights 
$\tilde{N}_{\gamma}$ was strongly dependent upon
the impact parameter $R_{P}$ in addition to the primary energy $E$.
From a detailed study of
MC simulated proton shower events generated with
CORSIKA,
we obtained a correction function to estimate the observed primary energy  
$\tilde{E}$ as a function of $\tilde{N}_{\gamma}$ and 
the long axis length $\tilde{L}$
calculated using Hillas analysis 
\citep{Hillas85}, where $\tilde{L}$ was used as an estimator of $R_{P}$. 
As a result, the total MC reconstructed energy resolution was estimated
to be 62\% 
by evaluating the rms of the $\Delta E/E = (\tilde{E} - E) / E$ distribution, 
for which the error was dominated by the ambiguity in $R_{P}$. 
The same reconstruction procedure was applied
both to the observed data and the MC data.
The observed and MC 
cosmic-ray flux spectra are shown in 
Fig.~\ref{fig:pedist}, 
in which the MC prediction
used the typically observed cosmic-ray flux with $E>$3~PeV 
given by
$ dN/dE = (3.7 \pm 1.1) \times 10^{6} E^{-3.0}
     ({\rm m}^2 {\rm sr \, \rm s \, GeV})^{-1}
$ fitting to 
the cosmic-ray flux observations in the knee region 
\citep{KASCADE05, TIBET08} 
with the power-law indices
$-$2.7 ($E<$3~PeV) 
and $-$3.0 ($E>$3~PeV) 
assuming continuity at the knee point.
We quote systematic uncertainties of 30\%,
mainly due to cosmic-ray flux spectrum observations around 
the knee (20\%--28\%),  
and partly due to their parameterization ($\sim$14\%). 
Since
the primary cosmic-ray components are observationally undefined, 
we present
the MC prediction of cosmic-ray flux spectra, assuming 
either only protons or irons as the primary cosmic rays 
in Fig.~\ref{fig:pedist}. 
In both cases,
the observed data and the MC prediction agreed well
on the normalization and the shape of the distribution
within the expected errors.
The estimation of the detection sensitivity of the Ashra light collector 
and the validity of the reconstruction procedure
were well demonstrated. 

For the $\nu_{\tau}$ search,
we used image data acquired using the trigger
for 197.1~hr 
in only case of the data status defined as good
out of the total observation time of 215.8~hr.
Using a Source Extractor
\citep{SExtractor96}, 
we extracted 
``clusters'' by requiring that 
of fired pixels ($N_{px}$) 
assigned to the 
cluster satisfied
$N_{px} >$4, 
where accumulated charge 
($Q_{px}$)
in the fired pixel ($px$) 
was required to be
$Q_{px} > $6~ADC units.
The primary cluster $C$, defined
as the cluster with the largest 
accumulated charge 
of fired pixels in the cluster
($\Sigma_{C}$),
was chosen as 
a pre-selected shower candidate for the event. 
To ensure that $C$
was within 
the effective fiducial FOV, 
the coordinates of the center of gravity 
(the center of the FOV was (0,0))
($X_{C},Y_{C}$)
of $C$ 
were required to satisfy 
$|X_{C}| < 13^{\circ}\!\!.8$, 
and 
($X_{C},Y_{C}$)
could not overlap any obstacles such as
rocks or the Mauna Loa surface within the
FOV.
Sometimes, ``car light events'' were triggered
when the light beams of a car  
on the Mauna Kea access road 
entered the FOV.
We determined the coordinates of the original position
($X_{\rm car},Y_{\rm car}$)
by averaging ($X_{C},Y_{C}$) 
for events that were apparently identified  
as car light events in the FOV.
The geometric distance 
$D_{\rm car}$
in the FOV between    
($X_{C},Y_{C}$) and 
($X_{\rm car},Y_{\rm car}$) 
had to satisfy
$D_{\rm car} > 0^{\circ}\!\!.2$. 
Remaining night sky backgrounds (NSBs) were rejected by
requiring that 
$\Sigma_{C} > 10\ \Sigma_{C'}$,
where $C'$ was defined as the second largest cluster.
The NSB rejection was also fairly effective 
at removing contamination from cosmic-ray $\mu$ events, in which 
Cherenkov lights from cosmic-ray $\mu$s passing through 
the input window glass (8~mm thick) of 
the photoelectric lens image tube
tended to separate into two or more clusters in the same image frame.
In the MC study,
the total selection efficiency 
for Cherenkov $\tau$ showers 
after satisfying all requirements
was estimated to be
99.9\%,
and
the expected number of events 
of cosmic-ray $\mu$ contamination 
was
5.8$\times 10^{-3}$
events.
We evaluated a small residual contamination of the final candidate samples
from secondary particles in large angle cosmic-ray air showers 
of 
1.3$\times 10^{-4}$
events
using CORSIKA 
with the curved earth option.
From the 197.1~hr of observation,
five clusters were pre-selected as shower candidates 
after the fiducial FOV cut.
Out of these five clusters, three were 
removed by the car light cut 
and two were removed by the NSB cut,
yielding a null result.

\section{Results}
\begin{figure}[bt!]
\begin{center}
\includegraphics[width=0.85\hsize]{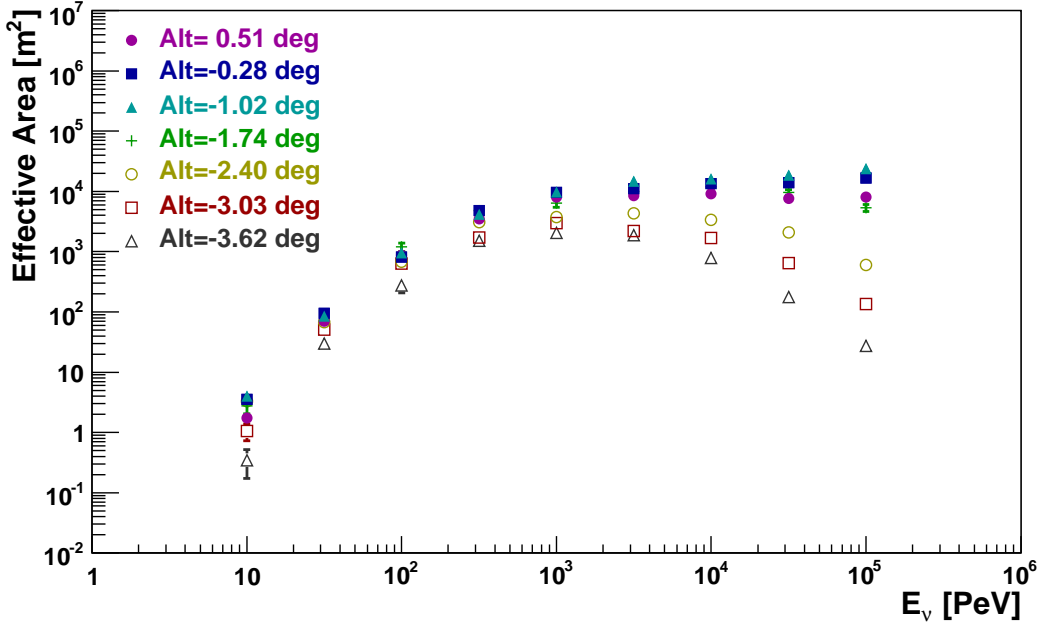} 
\end{center}
\caption{\label{fig:aper}
Effective area 
for Cherenkov $\tau$ showers induced by 
$\nu_{\tau}$s from the GRB081203A counterpart
as a function of $\nu_{\tau}$ energy ($E_{\nu_{\tau}}$)
assuming seven original positions on the counterpart trajectory 
(at elevation angles of $-3^{\circ}\!\!.62 \sim 0^{\circ}\!\!.51$) 
with the Ashra light collector.
}
\end{figure}
\begin{figure}[bt!]
\begin{center}
\includegraphics[width=0.85\hsize]{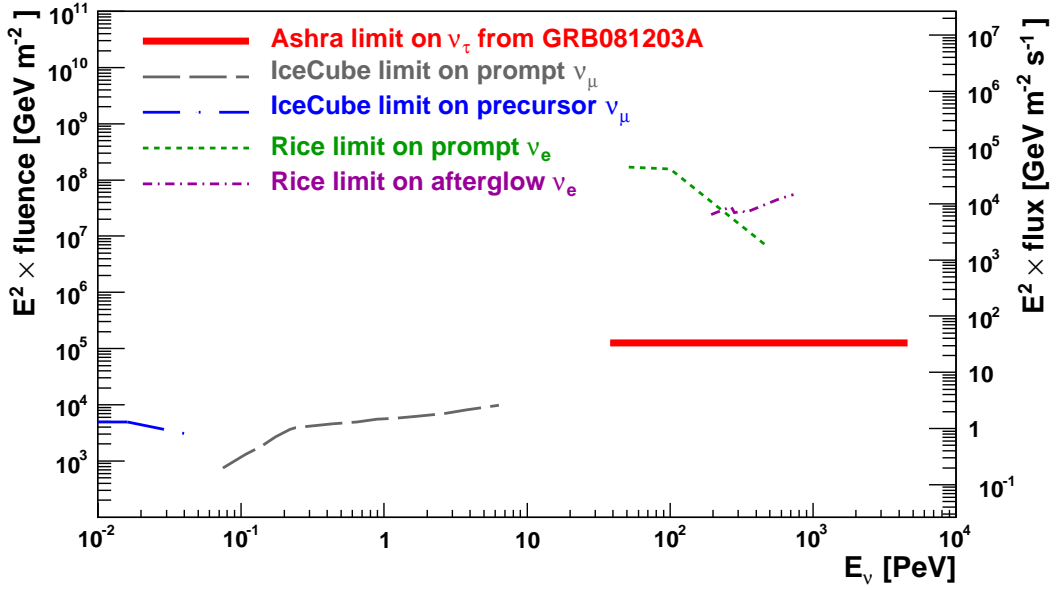} 
\end{center}
\caption{\label{fig:limit}
Ashra 90\%~CL limit 
(thick continuous line (red))
on $\nu_{\tau}$ fluence of precursor and afterglow emissions
from the GRB081203A counterpart
in the PeV--EeV region (see the text). 
For comparison,
IceCube limits 
\citep{IceCubeGRBs} 
in the prompt (long dashed dotted (gray)) and precursor (long dashed dotted (blue)) phases
and
RICE limits
\citep{RICEGRBs} 
in the prompt (dashed (green)) 
and afterglow (dashed dotted (magenta)) phases are shown.
The energy ranges
were defined 
as containing 90\%, 90\%, and 80\% of the expected signals from assumed spectrum 
for Ashra, IceCube, and RICE, respectively.
}
\end{figure}

Fig.~\ref{fig:aper} shows the effective areas 
for Cherenkov $\tau$ showers induced by 
$\nu_{\tau}$s from the GRB081203A counterpart
with changing $\nu_{\tau}$ energy ($E_{\nu_{\tau}}$)
as obtained from an MC study that 
assumed seven original positions on the GRB counterpart trajectory.
On the basis of the above null result and the estimated
effective areas, 
we placed 90\%~confidence level (CL) upper limits
on the $\nu_{\tau}$ fluence of precursor and afterglow emissions 
in the PeV--EeV region, 
for two 3780~s periods, the first between 2.83 and 1.78 hr before GRB081203A 
and the second between 21.2 and 22.2 hr after it,
as shown in Fig.~\ref{fig:limit}. 
The same fluence limits are set based on observations on the dates of
081122 (2430~s), 081124, 081125 (2000~s), 081126--081205, and 081206 (1840~s) 
for 3780~s each night unless otherwise noted.
As sources of systematic uncertainty in the MC estimates of the 
effective detection areas,
we considered the
$\nu_\tau$ charged current interaction cross section,
the energy loss by $\tau$ in the earth and the mountain, 
the geological model around the observation site,
and 
the gain calibration of the light collector,
which were evaluated to be
50\%, 50\%, 10\%, and 30\%, respectively
(Y. Asaoka \& M. Sasaki 2011, to be published). 
The conservatively combined systematic sensitivity error 
was obtained from the square sum of the above
uncertainties as 77\%, which affects the result of the
90\%~CL upper limit shown in Fig.~\ref{fig:limit},  
where
we assumed a typical $E_\nu^{-2}$ flux
to ensure unbiased constraints on
observationally undefined physical mechanisms of a GRB.

There are several models which predict VHE$\nu$ flux during our observation periods.
A precursor model based on existing supernova before a GRB
\citep{RMW04,Vietri98} 
allows VHE$\nu$ emission 
several hours before a GRB
from a choked jet 
generated by mass accretion. 
A cannonball (CB) model also predicts high-energy neutrino emission 
from a choked CB \citep{CBNeu01}.
Afterglow models based on external forward shock in a GRB
\citep{Vietri98a,Vietri98b,Vietri03,Li02}
and CB model \citep{CB_CR08,CB_IceCube} predict long-lived VHE$\nu$ emissions for 
more than a day after a GRB.
To evaluate quantitatively the afterglow flux 
in external forward shock
at 21.2--22.2~hr after the GRB,
the VHE$\nu$ flux expected from an afterglow model assuming interstellar medium in GRB030329 
\citep{RMW04} was rescaled to the flux from GRB081203A
by using isotropic gamma-ray luminosity ($L_\gamma^{\rm iso}$), redshift ($z$), 
and luminosity distance ($d_L$). 
$L_\gamma^{\rm iso} = 1.3 \times 10^{52}$~erg~s$^{-1}$ was calculated as an average luminosity 
during time period containing 50\% of the gamma-ray emission \citep{SwiftCat2010,GCN8599}.
$z=2.05$ was obtained from \citet{Grizm2009}.
Taking into account the expectation that VHE$\nu$ flux is proportional to optical afterglow 
flux \citep{Li02}, the VHE$\nu$ flux was corrected by the ratio of magnitudes (0.058)
measured after 1.5~hr from each GRB between GRB081203A \citep{GCN8604} 
and GRB030329 \citep{Prince03}.
The use of apparent magnitude corrected the effect of $d_L$ difference.
In addition, to account for decay of the optical light curve, 
the VHE$\nu$ flux was further corrected by the ratio of magnitudes ($4.8 \times 10^{-4}$)
between 21.7~hr \citep{GCN8615,GCN8618,GCN8619} and 294~s \citep{GCN8617} 
after the GRB081203A.
Finally, the redshift of neutrino energy was considered to obtain the neutrino spectrum.
Although the original model \citep{RMW04} calculated the VHE$\nu$ flux 
in reverse shock, it was applied to forward-shock afterglow by taking into account 
the optical-afterglow correction.
As a result, our limit between 21.2 and 22.2 hr after GRB081203A 
rejects a flux of $2 \times 10^5$ times larger than the model assuming neutrino oscillation.
If wind environment is assumed and the reverse shock model is applied with the 
correction due to $d_L$ \citep{LumDist} and $t^{-1}$ dependence of the expected 
neutrino luminosity \citep{Vietri98b},
the same limit rejects a flux of $3 \times 10^4$ times larger than the model.
Although our result is in agreement with the model, 
it might constrain some extreme cases such as very effective proton acceleration
and/or pion production in forward shock.
Due to the lack of detailed knowledge of the evolution
of the accelerated proton spectrum in the forward shocks,
there is huge ambiguity in calculation of neutrino fluxes \citep{Li02}. 
It is important to set observational limit
especially on each burst which would have a much different 
environment burst by burst.

For comparison,
Fig.~\ref{fig:limit} 
shows other observational limits on the $\nu$ fluence 
from point sources \citep{IceCubeGRBs, RICEGRBs}. 
Our results are the most stringent in the PeV--EeV region 
and complementary to the IceCube results for the sub-PeV energy region,
and indicate 
the advanced instantaneous sensitivity 
of the system 
even during this commissioning phase.
Full-scale Ashra observations are expected to
have 100 times better sensitivity,
since the trigger pixel size is halved (1/4 the pixel area) and 
the trigger sensor will cover the entire FOV of the light collector.
With this higher sensitivity,
contamination of 
0.55~clusters of air-shower secondary particles 
is expected in one year of observation data. 
The advanced angular determination of Ashra for
Cherenkov $\tau$ showers to within $0^{\circ}\!\!.1$ 
will allow perfect rejection against contamination from air-shower secondary particles, 
and will provide a viable search method for 
earth-skimming $\nu_{\tau}$ events, 
fully utilizing the zero background conditions. 
Our first search for PeV--EeV $\nu_{\tau}$ reported in this Letter 
complements other experiments 
in energy range and methodology, and suggests the prologue of ``multi-particle astronomy'' 
\citep{Sasaki00} with a precise determination of time and location.

We thank R.~Yamazaki for his comments. 
The Ashra Experiment is supported by  
the Coordination Fund for Promoting Science and Technology and  
by a Grant-in-Aid for Scientific Research from the 
Ministry of Education, Culture, Sports, Science and Technology in Japan. 


\begin{thebibliography}{48}
\expandafter\ifx\csname natexlab\endcsname\relax\def\natexlab#1{#1}\fi

\bibitem[{Abbasi {et~al.}(2010)}]{IceCubeGRBs}
Abbasi, R., {et~al.} 2010, \apj, 710, 346

\bibitem[{Abdo {et~al.}(2009)}]{Abdo09b}
Abdo, A.~A., {et~al.} 2009, \apj, 706, L138

\bibitem[{Abraham {et~al.}(2008)}]{Abraham08}
Abraham, J., {et~al.} 2008, \prl, 100, 211101

\bibitem[{Abramowicz \& Levy(1997)}]{ALLM}
Abramowicz, H., \& Levy, A. 1997, arXiv:hep-ph/9712415v2

\bibitem[{Ackermann {et~al.}(2010)}]{Ackermann10}
Ackermann, M., {et~al.} 2010, \apj, 716, 1178

\bibitem[{Agostinelli {et~al.}(2003)}]{Geant4}
Agostinelli, S., {et~al.} 2003, Nucl. Instrum. Methods Phys. Res. A, 506, 250

\bibitem[{Aita {et~al.}(2008{\natexlab{a}})}]{Aita07}
Aita, Y., {et~al.} 2008{\natexlab{a}}, Proc. 30th ICRC, 3, 1405

\bibitem[{Aita {et~al.}(2008{\natexlab{b}})}]{GCN8632}
---. 2008{\natexlab{b}}, GCN Circ., 8632

\bibitem[{Amenomori {et~al.}(2008)}]{TIBET08}
Amenomori, M., {et~al.} 2008, \apj, 678, 1165

\bibitem[{Andreev {et~al.}(2008)Andreev, Sergeev, Babina, \&
  Pozanenko}]{GCN8615}
Andreev, M., Sergeev, A., Babina, J., \& Pozanenko, A. 2008, GCN Circ., 8615

\bibitem[{Antoni {et~al.}(2005)}]{KASCADE05}
Antoni, T., {et~al.} 2005, Astropart. Phys., 24, 1

\bibitem[{Asaoka \& Sasaki(2011)}]{AsaokaSasaki11}
Asaoka, Y., \& Sasaki, M. 2011, Nucl. Instrum. Methods Phys. Res. A, 647, 34

\bibitem[{Bertin \& Arnouts(1996)}]{SExtractor96}
Bertin, E., \& Arnouts, S. 1996, A\&AS, 117, 393

\bibitem[{Besson {et~al.}(2007)Besson, Razzaque, Adams, \& Harris}]{RICEGRBs}
Besson, D., Razzaque, S., Adams, J., \& Harris, P. 2007, Astropart. Phys., 26,
  367

\bibitem[{Burlon {et~al.}(2008)Burlon, Ghirlanda, Ghisellini, Lazzati, Nava,
  Nardini, \& Celotti}]{Burlon08}
Burlon, D., Ghirlanda, G., Ghisellini, G., Lazzati, D., Nava, L., Nardini, M.,
  \& Celotti, A. 2008, \apjl, 685, L19

\bibitem[{Dar \& De~R{\' u}jula(2001)}]{CBNeu01}
Dar, A., \& De~R{\' u}jula, A. 2001, arXiv:astro-ph/0105094v1

\bibitem[{Dar \& De~R{\' u}jula(2008)}]{CB_CR08}
---. 2008, Phys. Rep., 466, 179

\bibitem[{De~Pasquale {et~al.}(2010)}]{DePasquale10}
De~Pasquale, M., {et~al.} 2010, \apjl, 709, L146

\bibitem[{Fargion(2002)}]{Fargion02}
Fargion, D. 2002, \apj, 570, 909

\bibitem[{Heck {et~al.}(1998)Heck, Schatz, Thouw, Knapp, \&
  Capdevielle}]{CORSIKA}
Heck, D., Schatz, G., Thouw, T., Knapp, J., \& Capdevielle, J.~N. 1998, Report
  FZKA 6019

\bibitem[{Hillas(1985)}]{Hillas85}
Hillas, A.~M. 1985, Proc. 19th ICRC, 3, 445

\bibitem[{Iyer~Dutta {et~al.}(2001)Iyer~Dutta, Reno, Sarcevic, \&
  Seckel}]{Dutta2001}
Iyer~Dutta, S., Reno, M.~H., Sarcevic, I., \& Seckel, D. 2001, \prd, 63, 094020

\bibitem[{Jadach {et~al.}(1993)Jadach, W{\c a}s, Decker, \& K{\"
  u}hn}]{TAUOLA24}
Jadach, S., W{\c a}s, Z., Decker, R., \& K{\" u}hn, J.~H. 1993, Comput. Phys.
  Commun., 76, 361

\bibitem[{Kachanov {et~al.}(1992)}]{YAP92}
Kachanov, V.~A., {et~al.} 1992, Nucl. Instrum. Methods Phys. Res. A, 314, 215

\bibitem[{Kann {et~al.}(2010)}]{SwiftCat2010}
Kann, D.~A., {et~al.} 2010, \apj, 720, 1513

\bibitem[{Krings(2010)}]{CB_IceCube}
Krings, T. 2010, PhD thesis, RWTH Aachen University

\bibitem[{Kuin {et~al.}(2009)}]{Grizm2009}
Kuin, N.~P.~M., {et~al.} 2009, \mnras, 395, L21

\bibitem[{Li {et~al.}(2002)Li, Dai, \& Lu}]{Li02}
Li, Z., Dai, Z.~G., \& Lu, T. 2002, A\&A, 396, 303

\bibitem[{Liu {et~al.}(2008)}]{GCN8618}
Liu, H., {et~al.} 2008, GCN Circ., 8618

\bibitem[{M\'es\'zaros(2006)}]{Meszaros06}
M\'es\'zaros, P. 2006, Rep. Prog. Phys., 69, 2259

\bibitem[{Mooney {et~al.}(1998)Mooney, Laske, \& Masters}]{Mooney98}
Mooney, W.~D., Laske, G., \& Masters, T.~G. 1998, J.~Geophys.~Res.~, 103, 727

\bibitem[{Mori {et~al.}(2008)}]{GCN8619}
Mori, Y.~A., {et~al.} 2008, GCN Circ., 8619

\bibitem[{Nousek {et~al.}(2006)}]{Nousek06}
Nousek, J.~A., {et~al.} 2006, \apj, 642, 389

\bibitem[{Parsons {et~al.}(2008)}]{GCN8595}
Parsons, A.~M., {et~al.} 2008, GCN Circ., 8595

\bibitem[{Prince {et~al.}(2003)}]{Prince03}
Prince, P.~A., {et~al.} 2003, Nature, 423, 844

\bibitem[{Razzaque {et~al.}(2004)Razzaque, M\'es\'zaros, \& Waxman}]{RMW04}
Razzaque, S., M\'es\'zaros, P., \& Waxman, E. 2004, \prd, 69, 023001

\bibitem[{Sasaki(2000)}]{Sasaki00}
Sasaki, M. 2000, in Proc. ICRR2000 Satellite Symposium: Workshop of
  Comprehensive Study of the High Energy Universe, ed. T.~Kifune {et~al.}
  (Kashiwa: ICRR, the University of Tokyo), 110

\bibitem[{Sasaki(2008)}]{Sasaki08}
Sasaki, M. 2008, J. Phys. Soc. Japan, 77SB, 83

\bibitem[{Sasaki {et~al.}(2008)}]{Sasaki07}
Sasaki, M., {et~al.} 2008, Proc. 30th ICRC, 3, 1559

\bibitem[{Sj{\" o}strand {et~al.}(2001)Sj{\" o}strand, Ed{\' e}n, Friberg, L{\"
  o}nnblad, Miu, Mrenna, \& Norrbin}]{PYTHIA6154}
Sj{\" o}strand, T., Ed{\' e}n, P., Friberg, C., L{\" o}nnblad, L., Miu, G.,
  Mrenna, S., \& Norrbin, E. 2001, Comput. Phys. Commun., 135, 238

\bibitem[{Ukwatta {et~al.}(2008)}]{GCN8599}
Ukwatta, T., {et~al.} 2008, GCN Circ., 8599

\bibitem[{Vietri(1998{\natexlab{a}})}]{Vietri98a}
Vietri, M. 1998{\natexlab{a}}, \prl, 80, 3690

\bibitem[{Vietri(1998{\natexlab{b}})}]{Vietri98b}
---. 1998{\natexlab{b}}, \apj, 507, 40

\bibitem[{Vietri {et~al.}(2003)Vietri, De~Marco, \& Gueta}]{Vietri03}
Vietri, M., De~Marco, D., \& Gueta, D. 2003, \apj, 592, 378

\bibitem[{Vietri \& Stella(1998)}]{Vietri98}
Vietri, M., \& Stella, L. 1998, \apj, 507, L45

\bibitem[{Volkov(2008)}]{GCN8604}
Volkov, I. 2008, GCN Circ., 8604

\bibitem[{West {et~al.}(2008)}]{GCN8617}
West, J., {et~al.} 2008, GCN Circ., 8617

\bibitem[{Wright(2006)}]{LumDist}
Wright, E.~L. 2006, \pasp, 118, 1711

\end{thebibliography}
\end{document}